\documentclass{PoS}

\def\beq{\begin{equation}}
\def\eeq{\end{equation}}
\def\beqa{\begin{eqnarray}}
\def\eeqa{\end{eqnarray}}

\title{Top quark pair and single top production at Tevatron and LHC energies}

\ShortTitle{Top quark pair and single top production at Tevatron and LHC energies}

\author{\speaker{Nikolaos Kidonakis}%
\thanks{This work was supported by the National Science Foundation under 
Grant No. PHY 0855421.}\\
        Kennesaw State University, USA\\
        E-mail: \email{nkidonak@kennesaw.edu}}

\abstract{I present the latest calculations  
of total and differential cross sections for top-antitop pair production 
and single top quark 
production via all main partonic channels. Higher-order corrections from 
the resummation of soft gluons are added through NNLL accuracy. 
Detailed numerical results are presented for approximate NNLO cross 
sections and top quark transverse momentum distributions at the Tevatron and LHC colliders.}

\FullConference{35th International Conference of High Energy Physics\\ 
		July 22-28, 2010\\
		Paris, France}

\begin{document}

\section{Top quark production and NNLL resummation}

Top quarks can be produced at hadron colliders via top-antitop pair 
\cite{CDFD0ttb} and single top \cite{D0CDFst} production channels. 
For $t{\bar t}$ production the 
leading-order (LO) partonic processes are $q{\bar q} \rightarrow t {\bar t}$, 
which is dominant at Tevatron energies, and $gg \rightarrow t {\bar t}$, 
dominant at LHC energies. For single top quark production the corresponding 
processes are  $qb \rightarrow q' t$ and ${\bar q} b \rightarrow {\bar q}' t$
($t$-channel), dominant at both Tevatron and LHC energies; 
$q{\bar q}' \rightarrow {\bar b} t$ ($s$-channel), 
which is small at both Tevatron and LHC; and 
associated $tW$ production, $bg \rightarrow tW^-$, which is 
very small at the Tevatron but significant at the LHC.
A related process is $bg \rightarrow tH^-$.

QCD corrections are significant for both $t{\bar t}$ and single top 
production. Higher-order corrections from threshold resummation 
of soft-gluon contributions further enhance the total cross section and 
top quark differential distributions \cite{NKRV,NKst,NKtH}. 
Recently these corrections have been 
resummed to next-to-next-to-leading logarithm (NNLL) accuracy, 
involving two-loop calculations of the soft anomalous dimensions. 
Approximate next-to-next-to-leading order (NNLO) total and differential 
cross sections have been derived from the NNLL resummed expressions 
\cite{NK2l,NKstWH,NKnnllttb}.
Below I show numerical results for the 
$t{\bar t}$ cross section and  the top quark $p_ T$ distribution at Tevatron 
and LHC energies, and for single top production via $s$-channel 
or via associated production with a $W^-$ or $H^-$ at Tevatron 
and LHC energies \cite{NKstWH,NKnnllttb}.

\section{$t{\bar t}$ cross section and top quark $p_T$ distribution at the Tevatron and the LHC}

\begin{figure}
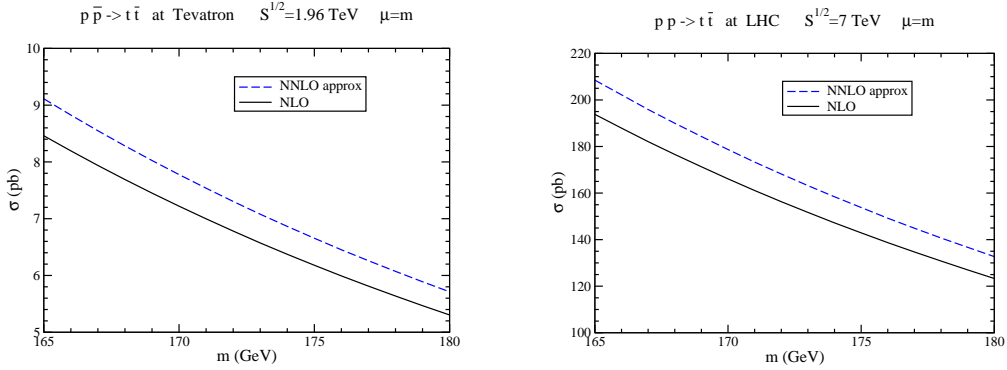

\begin{center}
\includegraphics[width=0.4\textwidth]{toptevplot.eps}
\hspace{10mm}
\includegraphics[width=0.4\textwidth]{top7lhcplot.eps}
\caption{Top-antitop pair cross section at the Tevatron (left) and 
the LHC (right).}
\end{center}
\end{figure}

We first study the total cross section for $t{\bar t}$ production (Fig. 1).
We derive an approximate NNLO cross section from the expansion of the NNLL resummed cross section. 
Using the MSTW2008 NNLO pdf \cite{MSTW2008}, we find at Tevatron and LHC 
energies 
\beqa
\hspace{2cm}\sigma^{\rm NNLOapprox}_{t{\bar t}}(m_t=173 \, {\rm GeV}, \, 1.96\, {\rm TeV})&=&7.08 {}^{+0.00}_{-0.32} {}^{+0.36}_{-0.27} \; {\rm pb} \, , 
\nonumber
\eeqa
\beqa
\hspace{2cm} \sigma^{\rm NNLOapprox}_{t{\bar t}}(m_t=173\, {\rm GeV}, \, 7\, {\rm TeV})&=&163 {}^{+4}_{-8}  {}^{+9}_{-9} \; {\rm pb} \, .
\nonumber 
\eeqa
At 14 TeV LHC collisions, we find $920 {}^{+50}_{-45}{}^{+33}_{-35}$ pb.
The first uncertainty is from scale variation by a factor of 2 around $\mu=m_t$
while the second is from the pdf \cite{MSTW2008} uncertainties.

\begin{figure}
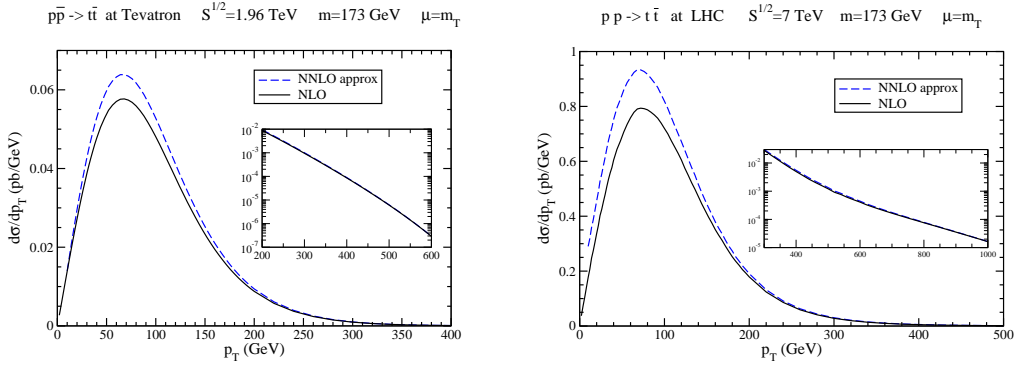

\begin{center}
\includegraphics[width=0.4\textwidth]{pttevplot.eps}
\hspace{8mm}
\includegraphics[width=0.42\textwidth]{pt7lhcplot.eps}
\caption{Top quark $p_T$ distribution at the Tevatron (left) and 
the LHC (right).}
\end{center}
\end{figure}

The top quark transverse momentum distribution at the Tevatron and LHC 
is shown in Fig. 2.

\section{Single top quark production: $s$-channel and associated production}

\begin{figure}
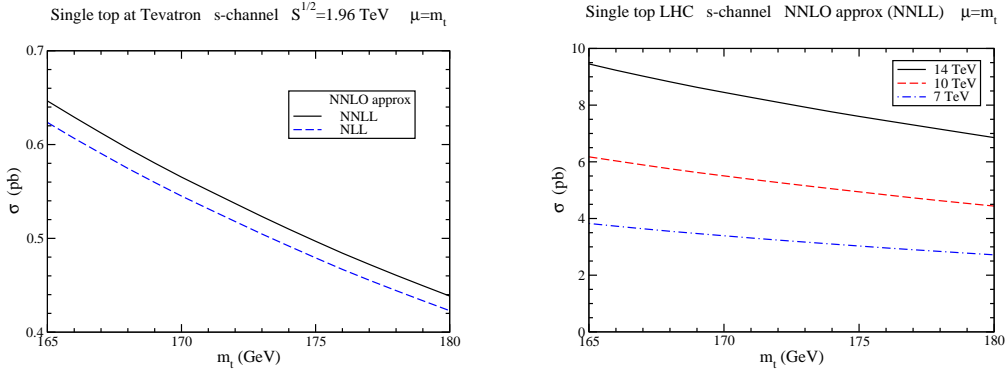

\begin{center}
\includegraphics[width=0.4\textwidth]{tevschmtplot.eps}
\hspace{10mm}
\includegraphics[width=0.4\textwidth]{lhctopschmtplot.eps}
\caption{$s$-channel single top cross section at the Tevatron (left) 
and LHC (right).}
\end{center}
\end{figure}

We continue with the $s$-channel single top cross section at the Tevatron 
(Fig. 3, left). We find 
\beqa
\hspace{2cm}\sigma^{\rm NNLOapprox,\, top}_{s-{\rm channel}}(m_t=173 \, {\rm GeV}, \, 1.96\, {\rm TeV})&=&0.523 {}^{+0.001}_{-0.005} {}^{+0.030}_{-0.028} \; {\rm pb} \, .
\nonumber
\eeqa
The cross section for single $s$-channel anti-top production at the Tevatron is identical.

The single top production cross section at the LHC in the $s$-channel 
(Fig. 3, right plot) is
\beqa
\hspace{2cm} \sigma^{\rm NNLOapprox,\, top}_{s-{\rm channel}}(m_t=173\, {\rm GeV}, \, 7\, {\rm TeV})&=&3.17 \pm 0.06 {}^{+0.13}_{-0.10} \; {\rm pb} \, .
\nonumber 
\eeqa
At 14 TeV, the result is $7.93 \pm 0.14 {}^{+0.31}_{-0.28}$.

For single antitop production at the LHC in the $s$-channel
we find $1.42 \pm 0.01 {}^{+0.06}_{-0.07}$ pb at 7 TeV; and 
$3.99 \pm 0.05 {}^{+0.14}_{-0.21}$ pb at 14 TeV.

\begin{figure}
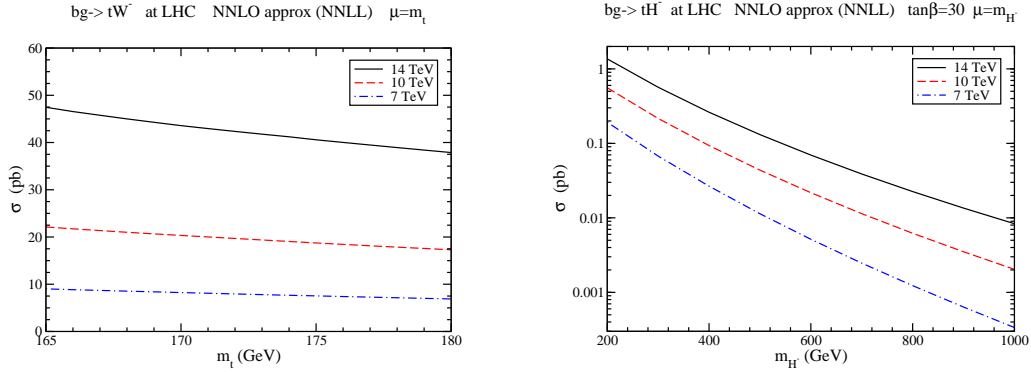

\begin{center}
\includegraphics[width=0.4\textwidth]{lhctWmtplot.eps}
\hspace{10mm}
\includegraphics[width=0.42\textwidth]{chiggsplot.eps}
\caption{$tW^-$ (left) and $tH^-$ (right) cross sections at the LHC.}
\end{center}
\end{figure}

The cross section for $tW^-$ production at the LHC (see Fig. 4, left) is 
\beqa
\hspace{2cm} \sigma^{\rm NNLOapprox}_{tW}(m_t=173 \, {\rm GeV}, \, 7\, {\rm TeV})&=&7.8 \pm 0.2 {}^{+0.5}_{-0.6} \; {\rm pb} \, .
\nonumber 
\eeqa
At 14 TeV, we have $41.8 \pm 1.0 {}^{+1.5}_{-2.4}$ pb.
The NNLO approximate corrections increase the NLO cross section by $\sim 8$\%.
The cross section for ${\bar t}W$ production is identical.

For $tH^-$ production (Fig. 4, right) 
the NNLO approximate corrections increase the NLO 
cross section by $\sim 15$ to $\sim 20$\%.

\end{document}